\documentclass[usenatbib]{mn2e}
\usepackage{epsfig}

\begin{document}
\title[A constant dark matter  halo surface density in galaxies]
      {A constant dark matter  halo surface density in galaxies}
      
\author [Donato et al.]
{F. Donato$^{1}$\thanks{E-mail:donato@to.infn.it}, 
G. Gentile$^{2,3}$, 
P. Salucci$^{4}$, 
C. Frigerio Martins$^{5}$, 
M. I. Wilkinson$^{6}$,
\newauthor{G. Gilmore$^7$,
~E.~K. Grebel$^{8}$,
~A. Koch$^{9}$,  
R. Wyse$^{10}$
}\\\\
$^{1}$Universit\`a degli Studi di Torino and INFN, via Giuria 1, 10125 Torino, Italy\\
$^{2}$Institut d'Astronomie et d'Astrophysique, Universit\'e Libre de Bruxelles, CP 226, Boulevard du Triomphe, B-1050 Bruxelles, Belgium \\ 
$^{3}$Sterrenkundig Observatorium, Ghent University, Krijgslaan 281, S9, B-9000 Ghent, Belgium \\
$^{4}$SISSA, Department of Astrophysics, via Beirut, 2-4, 34014 Trieste, Italy \\ 
$^{5}$Universidade Federal do ABC.Rua Catequese 242. 09090-400. Santo Andr\'e, Brasil  \\
$^6$Department of Physics and Astronomy, University of Leicester,  University Road, Leicester LE1 7RH, UK \\
$^7$Institute of Astronomy, University of Cambridge, Madingley Road, Cambridge, CB3 OHA, UK \\
$^8$Astronomisches Rechen-Institut, Zentrum f\"{u}r Astronomie der Universit\"{a}t Heidelberg, M\"{o}nchhofstr. 12-14, D-69120 Heidelberg, Germany \\
$^9$UCLA, Department of Physics and Astronomy,430 Portola Plaza, Los Angeles, CA 90095-1547, USA \\
$^{10}$Johns Hopkins University, 366 Bloomberg Center, 3400 North Charles Street, Baltimore, MD 21218, USA \\}

\maketitle 

\begin{abstract}
We confirm and extend the recent finding that  the central surface density  $\mu_{0D}\equiv r_0 \rho_0$ of
galaxy dark matter halos,  where $r_0$ and  $\rho_0$ are  the halo  core radius and central density, is
nearly constant and  independent of galaxy luminosity. Based on the co-added rotation curves of $\sim 1000$
spiral galaxies, mass models of individual dwarf irregular 
and spiral galaxies of late and early types  with
high-quality rotation curves and, galaxy-galaxy weak lensing signals from a sample of spiral and elliptical
galaxies, we find  that   $ \log \mu_{0D} = 2.15 \pm 0.2$, in units of $\log$(M$_{\odot}$
pc$^{-2}$).  We also show that the observed kinematics of Local Group dwarf spheroidal galaxies,  are
consistent with this value. Our results are obtained for galactic systems spanning over 14  magnitudes, 
belonging to different Hubble Types, and whose mass profiles have been determined by  several independent 
methods. In the same objects, the approximate constancy of $\mu_{0D}$ is in sharp contrast to  the
systematical  variations,    by several orders of magnitude,  of galaxy properties, including $\rho_0$  and
central  stellar surface density.
\end{abstract}


\section{Introduction}
\label{sec:introduction}
 
It has been known for several decades that the kinematics of disk galaxies exhibit a  mass
discrepancy (e.g. Bosma, 1978; Bosma 
\& van der Kruit,  1979;   Rubin, Thonnard \& Ford, 1980). More precisely, spirals  show  an inner
baryon dominance region   (e.g. Athanassoula et al. 1987,  Persic \& Salucci, 1988,
 Palunas \& Williams 2000), whose
size ranges   between  1 and 3 disk exponential lengthscales  according to the galaxy
luminosity (Salucci and Persic, 1999),  inside which  the observed  ordinary baryonic matter
accounts for the  rotation curve, but  outside which,   the distribution of the  baryonic
components  cannot justify the observed   {\it profiles} and sometimes    {\it amplitudes}
of the  measured circular velocities   (Bosma 1981, see also  Gentile et al. 2007).
This  is usually solved by adding an extra mass component, the dark matter (DM) halo. 
Rotation Curves (RCs) have been used to assess the existence, the amount and the
distribution of this dark component. Recent debate in the literature has focused on the
"cuspiness" of the dark matter density profile in the centers of galaxy halos that emerges
in Cold Dark Matter (CDM) simulations of structure formation (Navarro, Frenk \& White 1996, 
NFW hereafter;  Moore et al., 1999; Navarro et al., 2004; Neto et al. 2007) but is not seen
in observed data (e.g. de Blok, McGaugh \& Rubin, 2001; de Blok \& Bosma, 2002; Marchesini
et al., 2002; Gentile et al., 2004, 2005, 2007a), as well as on the various systematics of
the DM distribution (see Salucci et al., 2007). 

An intriguing general property of dark matter haloes was noted by
Kormendy \& Freeman (2004,  proceedings of IAU meeting), based on
halo parameters obtained by mass modelling   55 spiral
galaxy rotation curves  within the framework of the  Maximum Disk Hypothesis (MDH),
 whose  validity has been much debated 
 (Bosma 2004, Palunas and Williams 2000, Salucci and Persic 1999).	  
 Among other relations between the halo
parameters, they found that the quantity $\mu_{0D}\equiv \rho_0 r_0$,
proportional to the halo central surface density for any  cored halo distributions,
 is nearly independent
of the galaxy blue magnitude. Here $\rho_0$ and $r_0$ are,
respectively, the central density and core radius of the adopted pseudo-isothermal  cored dark matter
density profile $\rho(r)= \rho_0  r_0^2/(r^2+r_0^2)$.
 In particular, they found that this quantity takes a
value of $\sim 100 \, \rm M_{\odot} pc^{-2}$.  The Kormendy and Freeman analysis   
relies on the MDH, which fixes the value of  the  disk mass at its   maximum   compatible
with the observed rotation curve,  under the reasonable  hypothesis that mass follows light in the disk and  
 that the halo is not  hollow.  From the value of the disk mass RC fitting yield   the values of the two structural DM  parameters (i.e. $r_0$ and
$\rho_0$). As matter of fact,  MDH  allows to  {\it uniquely}   decompose the RCs - also those   that,   in term
of extension, spatial resolution, r.m.s errors,  non-axisymmetric motions,  cannot be  {\it
successfully}  analyzed  by  $\chi^2$ method assuming mass models with also the disk mass as a free
parameter.  The  MDH, on the other hand, may strongly bias  the determination of the  halo properties in the case in which  stars do not  dominate the inner parts of a galaxy.

More recently, Spano et al. (2008)  $\chi^2$ fitted  the RCs of 36 spiral
galaxies  by using  a mass model with a stellar disk and  a  cored dark sphere of density
\begin{equation}
\rho(r) = \frac{\rho_0}{\left(1+\left(\frac{r}{r_0}\right)^2\right)^{3/2}}. 
\label{rho}
\end{equation}
The  R-band surface brightness, via  the assumption of a 
constant mass-to-light ratio for the stellar component,   provided  the {\it profile}  of the stellar contribution to the circular velocity. They  showed that
\begin{equation}
{\rm log} \frac{\mu_{0D}}{\rm M_{\odot} pc^{-2}} = 2.2 \pm 0.25                                                  
\label{spano}
\end{equation}
or $\mu_{0D}= 150^{+100}_{-70} \, \rm M_{\odot}
pc^{-2}$, consistent with the findings of Kormendy \& Freeman (2004).  

In this paper, we will investigate the $\mu_{0D}$ vs magnitude relationship   for objects
whose central densities and core radii  vary by several orders of magnitude. We aim to
investigate  the above galaxy relationship  by applying    a number of   unbiased techniques
of DM decomposition   to new large  samples of galaxies of different Hubble Type  and
magnitudes. Given the wide-ranging nature of the data and of the  mass modelling  involved
in the present investigation, there is very little likelihood of obtaining a false-positive
result due to systematic errors and biases  in the  analysis or in the data.
\begin{center} 
\begin{figure}   
 
\vskip -2.1truecm
\psfig{file=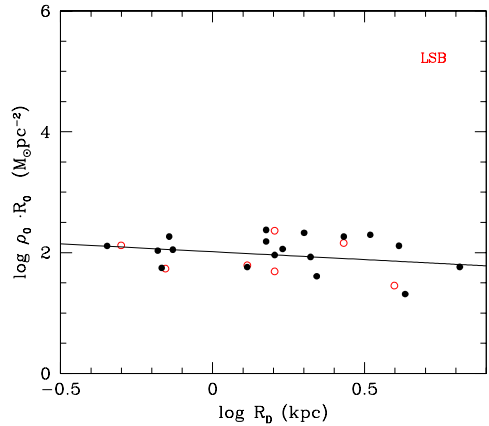,width=0.5\textwidth}   
\vskip -1.1truecm
\caption{The central halo surface density  
$\rho_0 r_0 $ as a function of disk scale-length $R_D$ for the Donato
et al. (2004) sample of galaxies. Open and filled circles refer to LSB
and HSB galaxies, respectively. The solid line is our best fit to the
data. }
\end{figure}   
\end{center}  

We will investigate: (a) a large sample of Spiral galaxies, analyzed by  $\chi^2$ fitting 
their Universal Rotation Curve (URC, PSS); (b) NGC3741, the most dark matter dominated
Spiral in the local Universe and DDO 47, a very well studied dwarf spiral (Gentile et al.
2005) by   $\chi^2$ modelling  their kinematics; (c) the THINGS  sample (Walter et al.
2008):  disk galaxies  with high quality RCs that have been mass modelled in {\it two}
independent ways,  1) by the standard  $\chi^2$ technique  and 2) by assuming the value of
the stellar disk mass from the galaxy color according to the prescription of
spectro-photometric galaxy models;  (d) a sample of Sa galaxies by  $\chi^2$ modelling 
their  kinematics; (e) a large sample of Spiral and Elliptical galaxies, by  $\chi^2$ 
mass-modelling the available weak-lensing shear measurements. We therefore  investigate 
Eq. \ref{spano} in a much wider range of  Hubble  types and  magnitudes and by exploiting a
larger number of techniques than previous works.  Finally, we   test the value of
$\mu_{0D}$  with the kinematics of six dwarf spheroidal satellite galaxies of the Milky Way
for which extensive stellar kinematic data sets are available.

In all cases  a cored dark matter halo provides a very satisfactory fit to the observed
data,   generally  superior to that  obtained  by assuming  a NFW profile for the DM  halo. 
The  success of the simple stellar disk + Burkert cored halo + HI disk model  in accounting for
the available kinematics   (both in absolute terms and with respect to different halo
models)  is  a strong support for  the  reliability of the derived halo  structural
parameters.  It is not an aim of this   paper to directly test the NFW halo  profile, and 
we will exclusively  work in the alternative  framework  of the cored halo profiles.  

With the exception of the weak lensing  analysis and  of dSph galaxies,  the  mass models   used
in this paper have been  obtained elsewhere,  in papers to which we redirect the reader  for
further information. 

In Section~2, we compute the quantity $\mu_{0D}$ for different families of galaxies, work
out its relation with galaxy magnitude.
A discussion of our result is given in Section 3.

\section {The $\rho_0 \MakeLowercase{r}_0$ vs magnitude  relationship}
\begin{figure*}   
\begin{center} 
\vskip -6.2truecm
\psfig{file=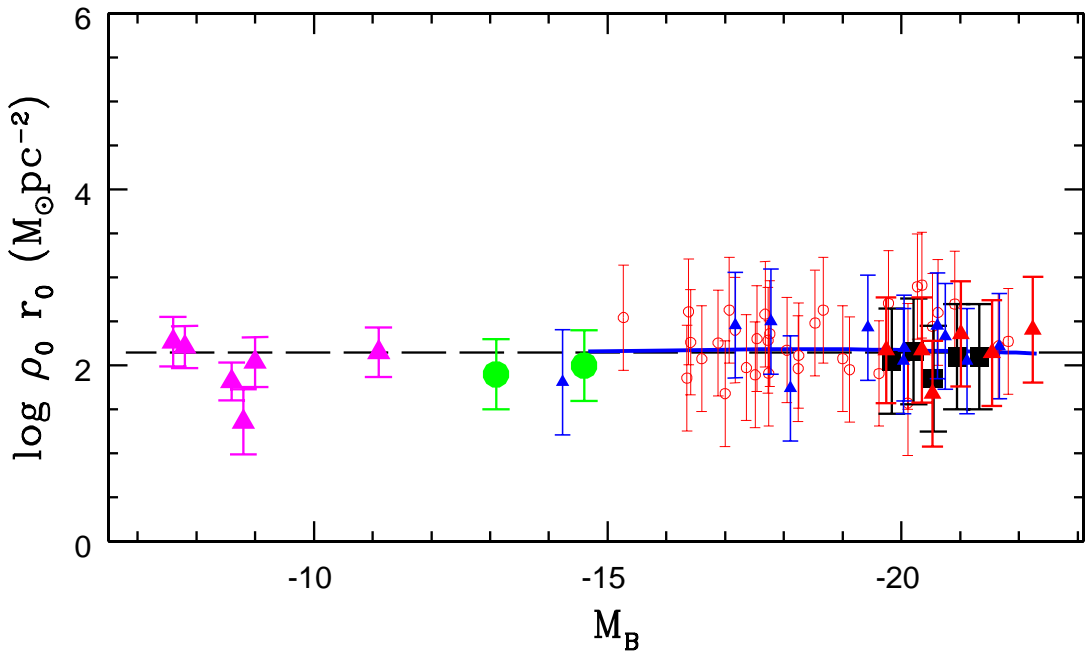,width=\textwidth}   
\end{center}
\vskip -1.9truecm
 \caption{ $\rho_0 r_0$ in units of $M_\odot$pc$^{-2}$ as a function of
galaxy magnitude for different galaxies and Hubble Types. The
original Spano et. al. (2008) data (empty small red circles) are
shown  as a  reference of previous work. The new results come from: the URC (solid blue line),
 the dwarf irregulars (full green circles) 
N 3741 ($M_B=-13.1$) and DDO 47 ($M_B=-14.6$),
    Spirals and Ellipticals investigated  by weak
lensing (black squares),  dSphs  (pink triangles), nearby spirals in   THINGS (small blue triangles), and   early-type spirals (full red triangles). The long dashed line is the result of this work. }
\end{figure*}   
 In this paper, we assume that the dark matter halo in each galaxy
follows the Burkert profile (Burkert 1995):
\begin{equation}
\rho (r)={\rho_0\, r_0^3 \over (r+r_0)\,(r^2+r_0^2)} .
\label{rho_bur}
\end{equation}
This profile, when combined with the appropriate baryonic gaseous and
stellar components, is found to   reproduce very well
the available kinematics of disk systems  (Gentile et
al., 2004; Salucci et al., 2003; Salucci \& Burkert, 2000; see Gentile
et al., 2007a for the case of the most extended RC). Moreover, it leads  to estimates of the
disk mass in good agreement with the  expectations from
stellar population synthesis models (e.g. Salucci, Yegorova \& Drory
2008, Gentile et al. 2004, Spano et al. 2008, see also Frigerio Martins
\& Salucci 2007).   
The  existence of a constant central surface density of dark
matter in  galaxies does not depend on which specific (cored)
density profile it is assumed  for the dark matter, i.e.  whether we adopt
any of the following: Spano et al. (2008; labeled as S hereafter),
Donato et al. (2004; D) or the present one (B). Different {\it
cored} mass models provide equally good fits to the same kinematical
data sets (e.g. Gentile et al., 2004).  All of them  can   describe  the actual  halo mass
profile $M_h(r, act)$ in the core region  by tuning the  values of the central density 
and  of core radius. The relations  $ M_h(r, act)=M_h(r, B)= M_h(r, S)= M_h(r, D)$ 
must hold, thus providing us with the  proportionality factors 
between the corresponding parameters (core radius and central density)  of
the different profiles. We find: $\log \mu_{0D}(B)= \log \mu_{0D}(D) +0.24 = \log \mu_{0D}(S) +0.1$. We use 
this small  corrections  to compare the values of $\mu_{0D}$  relative to  different  halo profiles. Of
course, at outer  radii --  outside the core  region and  often outside the last measured point --
each cored model has a  different well distinct velocity  behavior.  

One of the advantages of the  adopted Burkert  halo profile is that, at small radii and 
for  appropriate values 
of the parameter $r_0$, it  can   reproduce  the    NFW  velocity profiles to which, in any case,  it
converges for  $r >0.3  R_{vir}$.   Therefore,   with the  adopted   profile, the   RC data themselves  
discriminate,   by determining   the value of the    best fit parameter  $r_0$, the actual level of cuspiness
of the halo.  

Donato et al. (2004) analyzed the mass profiles of 25 spiral and LSB
galaxies obtained by $\chi ^2$ modelling their RCs.  The  successful 
 models had  cored dark matter halo profiles whose core radii correlated strongly
with the exponential disk scale length $R_D$ of their stellar
distributions. In Figure~1 we plot $\mu_{0D}$ as a function of $R_D$
for the Donato et al. (2004) sample. We see that the derived values
for $\mu_{0D}$ are almost constant, although $R_D$ varies by more than
one order of magnitude, consistently with the findings of Spano et
al. (2008) and Kormendy \& Freeman (2004). In addition, there is no
obvious difference between the results from High Surface Brightness
(HSB) galaxies and Low Surface Brightness (LSB) galaxies.  While this
result is in good agreement with Eq. \ref{spano}, it is important to
note that the two samples are similar, with five objects in common,
and the analysis employed is essentially the same.

Before adding   new  crucial   evidences  for a  relationship like Eq. \ref{spano},  
we would like to stress again that this  will   come from mass modelling techniques that are unbiased 
 towards any  particular DM profile,  and unable to artificially  create spurious  relationship between the DM mass parameters. 

It is useful to recall  the evidence from which we start (see Figure 2): 
 the relation found by  Spano et al. (2008)  for 36 spirals   and 
  the  above  $\mu_0D$ vs $R_D$ relationship  for the 25 spirals of the Donato et al. (2004) sample, both in qualitative  agreement  with Kormendy \& Freeman (2004).      

We now calculate the central surface density $\mu_{0D}$ for the family of  Spirals by means of their
Universal Rotation Curve (URC) (Persic, Salucci \& Stel, 1996, PSS hereafter). 
This curve, {\it on average}, reproduces well  the
RCs of late type (Sb-Im)  Spirals out to their virial radii $R_{vir}$ (PSS; Salucci et al., 2007). 
The URC is built from (a) the co-added 
kinematical data of a large number of Spirals (PSS; see also Catinella et al., 2006) and (b) the disk mass
versus halo virial mass relationship of  Shankar et al. (2006). By   $\chi^2$ fitting  the URC with a Burkert
halo + a Freeman disk velocity model,  with  no assumptions  on the amount of baryonic matter, we obtain
$\rho_0$ and $ r_0 $  as  the best-fit    values   (see equations 6a, 7 and 10 of Salucci et al., 2007). The
corresponding $\mu_{0D}$ values  are plotted vs. $M_B$  as a solid line in Figure~2. The URC, 
derived from co-added rotation curves of objects with the same  luminosity,    traces their  ensemble-averaged
gravitational potential. This  is extremely useful:   the  consequent mass model is free from the  
particularities (internal r.m.s.,   non-axisymmetric motions, observational errors) that  affect, at different levels,   almost every  {\it individual} rotation  curve  
 and the  ensuing mass model; this particularities, in fact,   get   averaged out in the URC  construction.

 On the other hand,  for the  task of determining the DM
structure parameters, the coadded  RCs are not  sufficient in that, at a fixed luminosity, there could be   a  Cosmic Variance around ``the average galaxy''. Then, in order to  assess  the  universality of Eq. \ref{spano},   we will investigate the DM mass structure in individual objects supplementing new observational  data to those of   Kormendy and Freeman 2004, Spano et al.  and of  Donato et al. (2004).

de Blok et al. (2008)  measured  high-resolution rotation curves for  a sample of late  spirals  belonging to  THINGS (The
HI Nearby Galaxy Survey). We select from this sample  the objects in which  the mass modelling yields  reliable estimates of the dark matter structural parameters.  We have rejected objects in which  a)  the kinematics  is clearly   affected by
non-circular motions, b) the stellar mass component  strongly dominates  the  galaxy
potential out  to the last measured point, preventing us from determining the  properties of the underlying  dark matter halo, or c) very different models are found to   equally fit the RC. 
With these selection criteria, we rejected 6 galaxies out of 17.
This selection, though  mandatory   to successfully probe the DM potential (e.g. Lake \& Feinswog 1989),   
limits  the investigation of the DM distribution of galaxies  by means of their kinematics and photometry. For instance we can use only   objects in which  a) {\it neatly both} dark component  and  the  stellar components {\it of known distribution}  affect (at different radii)  the available kinematics   b) non-circular motions are modest .

The    rotation curves  in the  Blok el al.'s THINGS sample were   modelled  with a spherical pseudo-isothermal dark halo plus
an  HI disk  and a   stellar disk whose free mass  parameters are obtained by $\chi^2$ fits. 
In addition to the standard method in which the  stellar mass-to-light ratio is a free
parameter, they also modelled their RC's by  assuming for the latter  quantity the values obtained from the
galaxy colors as  predicted by  spectro-photometric models with a  1) diet-Salpeter or a 2)  Kroupa IMF. For
each object we take their results in the following way: we  average  the value of  $\rho_0 r_0$ obtained by 
the latter  two methods, and then we  average the result with the value obtained by the un-constrained mass
model.  Notice that we take also  the values of $\rho_0 r_0$ coming from the  spectro-photometric method of mass modelling, although the latter  may be less accurate than the $\chi^2$ one (e.g. Salucci et al. 2008)  because we want an independent check on  the mass modelling procedure. In any case,   the values obtained by $\chi^2$ fits are within the shown errorbars.

 We found that the galaxies DDO  154, N 925, N 2366, N 2403, I  2574, N 2976, N
3198,  N 3621, N 5055,  N 6946,  N 7331 satisfy the  above discussed selection criteria. The resulting  mass models  well reproduces the  RCs   and the relative  halo parameters are  derived within a reasonable uncertainty ($\leq
50\%$).  The resulting values of $\mu_{0D}$  are plotted in Figure (2).

We extend the relationship down to   the   lowest luminosities of   disk systems by means of the  nearby
dwarf galaxy NGC 3741 ($M_B=-13.1 $): it represents the very numerous dwarf disk objects which are dark
matter dominated down to one disk length-scale or less, and in which the HI gaseous disk is the main baryonic
component.  In addition, this  galaxy has an extremely extended HI disk, which allowed
Gentile et al. (2007b) to carefully trace the RC and therefore its gravitational potential out to
unprecedented distances,  relative to the extent of the optical disk. The data probe out of 7 kpc
(equivalent to 42 B-band exponential scale lengths) with  several independent measures within the estimated
halo core radius. By standard  $\chi^2$  fitting,  the RC  was  decomposed  into its stellar, gaseous and
dark (Burkert halo)  components.  The resulting  best fit  mass model  very well reproduces the observed RC
(Gentile et al., 2007b): the corresponding $\mu_{0D}$ is plotted in Fig. 2 as a filled green circle. This result  
is seconded by  DDO 47, another faint dwarf spiral. Gentile et al. (2005) have  mass modelled  its RC    in 
the same  way as described above. We plot the relevant quantities   in Fig. 2 as another filled green circle, at
$M_B=-14.6$.     The relatively large error-bars of both estimates  is due to uncertainties in the distance,
that affects any nearby object, and not by the mass model itself, which is virtually free from the  uncertainties  in
the estimate of   the mass of the  stellar disk (which for these object is negligible). 

It is worth to notice that Burkert (1995), in his pioneering study on the DM structure in galaxies, for  an
handful of  dwarfs with absolute blue magnitudes ranging between -14.5 and -17.0 and modelling  their   low
spatial resolution  HI RCs, found values of $r_0$ and $\rho_0$ that lead,  in these objects,  to   $90  \leq  
\mu_{0D}/ ({\rm M}_{\odot} {\rm pc}^{-2}) \leq 140   $\,  in agreement with our results.

To  investigate the opposite end  of Hubble Spiral  Sequence,  i.e. the Sa galaxies,  disk systems embedded
in  a relevant spheroidal stellar component,  we resort to the   mass models that have become  
recently  available (Noordermeer 2006, Noordermeer et al. 2007). From this sample, using to the selection
criteria discussed above, we take the following galaxies:   
N 2487,  N 2916,   N 2953,  N 3546,   UGC 8699,   UGC 11852. 

We  reject 11 galaxies out of 17.  Notice that,  only  for a small fraction of the rejected objects in the  THINGS and Noordermeer sample  the failure of the  mass modelling is due to poor  kinematics.  In most of the cases,  it  originates from the presence of a strong inner dominance on the galaxy dynamics  of  {\it two }  baryonic components (the disk and the bulge), and it  may reflect  an intrinsically  complex inner mass distribution. On the other hand, systems with a  multi-component strong  central baryonic  mass concentration 
likely  underwent secular physical processes  that may  have affected   the original  distribution of  
the dark matter halo (Heller et al. 2007, Athanassoula, 2008) making them complex systems that  must be investigated more accurately by future 
studies.

The mass
models  are based on RC $\chi^2$  decompositions that  include  a  stellar bulge,  a stellar disk, a 
neutral gas disk and a  pseudo-isothermal (cored)  dark matter halo. 
The resulting  $\mu_{0D}$ are plotted in Fig. 2.

We now  derive  the  galaxy  mass distribution   by measuring     their  gravitational potential in   a
different way  from that  employed so far. This will test both  the observational data and the fundamental
assumptions underlying the results shown above. From  the galaxy-galaxy weak-lensing signals of a large
sample of Spiral and Elliptical galaxies,  we  determine their DM  distribution. The basic data is  the
azimuthally-averaged tangential shear $\gamma(r)$ recently measured   for a sample containing about $10^5$ isolated
objects split into 5 luminosity bins (Hoekstra et al. 2005),  as a function of the galactocentric radius. The
sample spans a good  luminosity range of Spirals, while the most luminous bin is likely dominated by the
biggest Ellipticals in the local Universe. Data extend  from $R_i$ =70 kpc  out to   $R_f $ =560  kpc from
the center of the lenses. In  this radial  range, the galaxy stellar component (a Freeman disk for 
spirals,   a Sersic spheroid for ellipticals)  contributes negligibly  to the shear: the spheroid half-
light radius does not  reach  $10$  kpc a distance  $ << R_i $. The  mass model, therefore,    includes only
a  (Burkert) dark  halo. Notice, however,  that,  while     we need  kinematical   data at radii  well  inside    $r_0$  to detect in a RC  a Burkert core (of size $r_0$), in the  tangential shear, instead,  the effect of a Burkert profile     extends further out,  up  to $2\  r_0 $, i.e. for the most luminous objects, it extends out to $\sim R_i$. The present weak lensing  data   are (marginally)  able to to measure the values of   $\rho_0$ and $r_0$. Formally these  are  obtained   by $\chi^2 $ modeling $\gamma(r) $ with a Burkert mass profile. The details are presented in Appendix A and the  resulting $\mu_{0D}$  are plotted in Fig. 2 (as
solid squares). Thus, we applied the same technique to the same kind of data  both for Spirals (all luminosity  bins but the last)   and for  Ellipticals (the last bin). We  found no difference in
the DM profile systematics and in particular  in the value of $\mu_{0D} $. 
Then, from our collection of values, at the level of 0.2 dex,  no substantial  differences emerge between the
values of $\mu_{0D}$ estimated from different types of data or between Spiral and Elliptical galaxies. It
thus appears that the central surface density of DM halos assumes a nearly constant value with respect to
galaxy luminosity, over a range of at least nine magnitudes.

For illustrative purposes, we compare our results with  those of  Spano et al (2008). 
We plot their data in Figure 2. Let us remark that their data are not included in our present  sample: indeed, 
because we want to raise  our claim in an independent way from their work, and their data are used as a 
consistency check. However,  we  remove two  objects with an  enormous  uncertainty (i.e.  $>$ than a factor 10 ) on the  best fit value  of one of the  two parameters  $r_0$, $\rho_0$ (private communication, UGC 3876 and  UGC 4456).  

\subsection{Milky Way satellites}

This result can be extended to lower magnitudes by means 
of the Milky Way satellite dwarf spheroidal (dSph) galaxies,
the smallest and most dark matter dominated systems known in the universe
(see e.g. Mateo, 1998; Gilmore et al. 2007, and references
therein). Their low HI gas content is another property that sets them apart as a
galaxy class (e.g. Grebel, Gallagher \& Harbeck, 2003). In a recent study of six
dSphs Gilmore et al. (2007) showed by $\chi^2$ techniques  that, 
assuming spherical symmetry and velocity isotropy, the stellar kinematics and
photometry of dSphs are consistent with their occupying cored DM haloes. Our
current lack of knowledge about the anisotropy of the stellar velocity
distribution, make their density profiles  not uniquely constrained by the data. 
Cusped models can also  reproduce the dispersion velocity data  in most dSphs
(Gilmore et al. 2007; Koch et al. 2007; Battaglia et al. 2008), modulus an
appropriate  run with radius of the anisotropy parameter.  
Bearing this caveat in
mind, we  will assume spherical symmetry and velocity isotropy in estimating $\mu_{0D}$. 
The observed  stellar density $\nu(r)$ distribution is  well represented by a  Plummer sphere:   $\nu(r) \propto (1+(r/a)^2)^{-5/2}$  with $a$ the half light radius.
This stellar spheroid  is tracer of  but a negligible  source for the gravitational
potential: its mass is only   $10^{-3}$ times the dark mass  inside $a$ (Gilmore et al., 2007).  
The full  mass modelling of these objects are given  in Salucci et al. (2009). Here we compute the relevant structural  parameters with a simplified approach. We  realize   that the 1-D stellar velocity  dispersion $\sigma(r)$   are radially  very slowly varying and we assume,  for the purpose of this work,  that is constant:   $\sigma(r)=\sigma_0$. 
Therefore, within the above assumptions, from the Jeans equation the halo mass can be computed by: $G^{-1} \frac {r^2}{\nu(r)} \frac{ d\nu(r)}{dr} \,  \sigma_0^2$ that leads to  $ 5 \, G^{-1}\frac{ r^3}{a^2
\left(\frac{r^2} {a^2}+1\right)} \, \, \sigma_0^2 $ with the values of $\sigma_0$ and $a$ given in Gilmore et al. (2007). The $r^3$ dependence at small radii indicates the presence of a core. Indeed, the above mass distribution  can be successfully fit by a Burkert profile with a  $ r_0\simeq  a$ and  $\rho_0 \simeq  2.7\, G^{-1}
\sigma_0^2/a^2$. The corresponding $\mu_{0D}$ are plotted in Figure~2 as
triangles with the error bars reflecting  the statistical errors in the estimation
of the parameters from the observed data.

 As a result, the values  $\mu_{0D}$ keep
constant around $\simeq$ 100 M$_{\odot}$ pc$^{-2} $  also for this sample of dwarf
galaxies. This outcome  is far from trivial.  In   dSphs both the  central halo density
and the core radius take much higher and much  smaller values with respect to
those of the  faintest  spirals, which are  objects 5 magnitudes brighter.
Such variations are nevertheless fine-tuned  so  that   the product   $\rho_0 r_0$
remains  almost constant,  despite the strong discontinuity of the two separate
quantities (and of any other  galaxy property).

Finally the "well noted curiosity" that all  dSph halos   contain roughly equal masses
interior to about 0.3-1.0\,kpc (Mateo {\em et al.} 1998; Gilmore {\em et al.}  2007;
Strigari et al. 2008) can be understood.   For a Burkert profile  the   constancy of
$\mu_{0D}$ implies the mass constancy inside any fixed physical radii and
viceversa.

\section{Results}

We have assembled  and discussed data on the DM halo mass distribution for many galactic systems of different
Hubble Type including dwarf disks and spheroidals, spirals, ellipticals, spanning almost the whole galaxy  
magnitude range  $-8 < M_B < -22$ and  gaseous-to-stellar mass fraction range  (wide as  many orders of
magnitude). The mass modeling of such objects has been  carried out by using different and independent
techniques,  none of them capable to bias  the resulting  DM  distribution towards an artificial
relationship.   

 Then, our current knowledge of the distribution of DM in dSphs   suggests that the relation
$\rho_0 r_0 \approx$ constant may extend to the  faintest galaxy  systems, and then we can
claim   valid over a range of fourteen magnitudes in luminosity and for all Hubble Types:
\begin{equation}
{\rm log} (\mu_{0D}/{\rm M_{\odot} pc^{-2}}) =  2.15 \pm 0.2  
\label{we_find}
\end{equation}
or  $\mu_{0D}= 140^{+80}_{-30} \,\, {\rm M_{\odot} \, \,  pc^{-2}}$
\begin{figure*}   
\begin{center} 
\vskip -.0truecm
\psfig{file=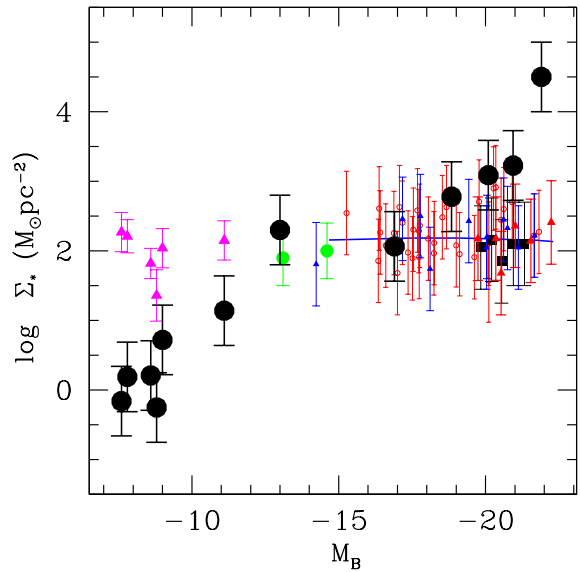,width=0.5\textwidth}   
\end{center}
\vskip -0.8truecm
\caption{Stellar central surface density $\Sigma_{*0}$ in units of
$M_\odot$pc$^{-2}$ (full black circles) as a function of
galaxy magnitude for different galaxies and Hubble Types. As comparison the values of $\mu_{0D}$
are also shown with the same coding of Fig 2.} 
\end{figure*}   
 
The observed galaxy  kinematics  are well reproduced  by a Burkert  cored halo profiles with two
structural parameters: a central halo density $\rho_0$ and a core
radius $r_0$, whose respective values span several orders of
magnitude: $ 6\times 10^{-23} {\rm g/cm^3}
\leq \rho_0 \leq 10^{-25} {\rm g/cm^3}$ and $ 0.3$ kpc $\leq r_0 \leq
30$ kpc.  In spite of dealing with spirals/ellipticals with such different
DM physical properties,  that  parallels the large systematical variations of 
properties of the luminous counterparts, we have found that their DM surface
densities $\mu_{0D}\equiv \rho_0 r_0$ remain  almost constant.
Our finding indicates that the DM central surface density in galaxies  is essentially
independent of their luminosity (mass).  

Our result crucially strengthens and enlarges the earlier findings  by Kormendy \& Freeman
(2004) and Spano et al. (2008) of a constant  ($\sim 100$  ${\rm M_{\odot} pc^{-2}}$) value
for the surface density  among  some classes of galaxies,  a result obtained by extracting DM halo parameters
from the galaxy kinematics of  relatively small  samples of galaxies  within, for  the
first case, an  assumed theoretical framework.   Eq. \ref{we_find}  relies on  a much larger number
of  objects across more Hubble-types and a much  wider luminosity range. Furthermore,  
they  are obtained  from mass modelling  performed by  {\it model independent} techniques of
both    {\it individual}   and {\it co-added} galaxy  kinematics/shear. While  the
URC/shear analysis have  provided  reliable estimates of the average value of $\mu_{0D}$ 
for galaxies  of a given luminosity,  the detailed studies of individual objects have
detected small the cosmic variance around this average.   

We cannot presently  exclude that  $\mu_{0D}$ has  systematical  or object  by object
variations at the level smaller than  $ 30 \%$ of its value,   neither  that Eq. \ref{we_find} be  a
byproduct of some more fundamental relationship, however,  we can claim that Fig. 2 and Eq.
\ref{we_find},  alongside with the support of previous work,  points  to an (unexpected)   DM  
property that it is not a spurious effect due  to   adopted  selection criteria, 
observational errors  and/or incorrect assumptions in the galaxy  modelling.

\section{The intriguing relation between $\mu_{0D}$ and the stellar central surface density}

The  constancy of $\mu_{0D}$ is particularly relevant also because in stark contrast to the
observed  variations of  {\it stellar}  central surface density $\Sigma_*$ of  galaxies of
different Hubble Type and magnitudes, i.e. of its luminous counterpart.
 $\Sigma_*$ (the details  on the following estimates can be found  in the  papers  cited above) shows a strong luminosity
dependence, as illustrated in Fig. 3. In Spirals, PSS   find that   $\Sigma_*$    
increases  with luminosity:   $\Sigma_*  \sim 800 \, {\rm M_\odot pc^{-2}}$ at about $M_B=
-22.5$ and   $ \Sigma_*   \sim 50 \, {\rm M_\odot pc^{-2}}$ at $M_B=-17$;  in dSphs, 
obtained by the central surface photometry and by assuming $M/L_V=1  $ $\Sigma_*$ takes
extremely low values: $ (1-10)  \, {\rm M_\odot pc^{-2}}$; when
computed in Ellipticals  by the central surface photometry and by assuming $M/L_B= 5$,  
it easily exceeds values of  $ 10000\, {\rm M_\odot pc^{-2}}$.  Given these very  large  variations  with galaxy
luminosity, the uncertainties  related to the  estimate of $\Sigma_*$, of the order of 30 \%,  are
irrelevant here.  We  can   draw the following   consequences: a) the central  surface
density is the only DM quantity which  is not correlated with its stellar  analogous, 
differently from  any other (core radius, central spatial density, mass etc.),  b) the
stellar component  dominates the center of all galaxies,  but the dwarfs where it is
surprisingly very sub-dominant.  

\section{Discussion and conclusions}

Let us  consider how the approximate constancy of
$\mu_{0D}$ with $M_B$ is related to the correlation between $r_0$ and
$\rho_0$,
\begin{equation}
\log r_0 = A  \log \rho_0 +C
\label{burkert}
\end{equation}  
found in spiral galaxies (e.g Burkert, 1995). Clearly, if $\mu_{0D}$ were exactly constant, this would imply
that $\rho_0 \propto r_0^{-1}$ and viceversa.   However,  the  variations  in Eq. \ref{we_find},   as well as
the observational uncertainties irrelevant  for the run of    $\mu_{0D}$  with luminosity, are substantial if
one   wants to  invert   relation  \ref{we_find} to    obtain a $\rho_0-r_0$ relation.  In fact, the  
propagated uncertainties from  Eq. \ref{we_find}  would make the estimate of  $r_0$  from $\rho_0$ uncertain  within a factor  not less than  $2\times 10^{0.2}$ and  occasionally as big as  $2\times 10^{0.5}$,  i.e.  useless   for
mass modelling aims. Furthermore, given the large range of  the values of  $\rho_0$ and $ r_0$ in galaxies, eq (4) cannot  make any  claim beyond to confirm   a general trend between the two structural halo quantities, with  $A \sim 1$.  The  relationship   between  $\rho_0$ and $ r_0$  must be worked out  separately from the study of Eq. \ref{we_find}, from a  properly selected observational data and with suitably performing   methods of mass
modelling. 
   
It is remarkable that the  constancy of $\mu_{0D}$ can be related to  well-known  
scaling laws of spirals. Let us define  $M_{\rm h0}$ and  $V_{\rm h0}$ is the enclosed 
halo  mass inside $r_0$ and the  halo circular velocity at $r_0$.
Since for a Burkert halo  $M_{\rm h0} \propto
\rho_0 r_0^3$,   then eq (4)  implies  $M_{\rm h0} \propto V_{\rm h0}^4$   which immediately reminds
a sort of Tully-Fisher relation (e.g. Freeman, 2004,  McGaugh 2005).

Moreover, we can  estimate the ratio between the contribution to the circular velocity  from the  disk and the dark halo at $r_0$. From $\mu_{0D}=const$ one has,  for a Burkert halo:  $V_{h0}  \propto r_0^{0.5}$.   By  means of  the relationship in eq(3) of   Tonini et al. (2006) that  relates  in Spirals $R_D$ with  $M_D$  and from  the relation $r_0 \propto R_{ D}^{1.05}$ (Donato et al. 2004),  one can compute the disk contribution $r_0$:    $V_{d0}  \propto r_0^{0.8}$.  From these dependencies we get that  the  velocity contributions  fraction  is proportional to $R_{D}^{-0.6} \propto L_B^{-0.3}$,  in good agreement with a  main scaling law of spirals (Persic and Salucci, 1988, PSS). The constancy of $\mu_{0D}$ seems therefore related to the fact that  less luminous objects have,  in proportion, more dark matter.

Considering that DM haloes are (almost)  spherical systems it is  surprising that their central surface
density plays  a  role in galaxy structure.  One could wonder whether the physics we witness in $\mu_{0D}$  is
instead   stored separately in the quantities $r_0$ and $\rho_0$.  This reasonable interpretation has 
however a problem:   $r_0$ and $\rho_0$ do  correlate with the luminous counterparts 
(the disk length-scale and stellar central surface density) while  $\mu_{0D}$ does not.

The evidence that the DM halo central surface density $\rho_0 r_0$
remains constant to within less than a factor of two over at least
nine (and possibly up to fourteen) galaxy magnitudes, and across
several Hubble types (we note, however, that for early-type spirals we have 
limited information),  obviously indicates  that this quantity  may 
hide an important physical meaning  in the DM distribution of galaxies. Presently this finding is
 surprising, as it is difficult to envisage how such a relation
can be achieved  across galaxies which range from
dark-matter-dominated to baryon-dominated in the inner regions. In
addition, these galaxies have experienced significantly different
evolutionary histories (e.g. numbers of mergers, significance of
baryon cooling, stellar feedback, etc.). 
 
Finally, let us spend a few words of caution about the result we claim in  this paper.  Further  investigation is still needed before that we can correctly frame it in a cosmological context.  In fact,   although the number of objects for which    a reliable DM mass distribution has been  obtained is impressive, it is still quite limited with respect to the cosmic variance of present day galaxies. Moreover,  some types of objects such as those with distorted kinematics or those in which a bi-component stellar distribution has a strong central concentration,  still escape a satisfactory analysis.

\section*{Acknowledgments}
We thank the referee, Albert Bosma, for useful comments which 
improved the quality of the paper.
CFM ackowledges support from the FAPESP/Brasil Fellowship. MIW
acknowledges support from a Royal Society University Research
Fellowship. GG is a
postdoctoral fellow with the National Science Fund (FWO-Vlaanderen).

\appendix
\section{}

\begin{figure*}
\begin{center}
\psfig{file=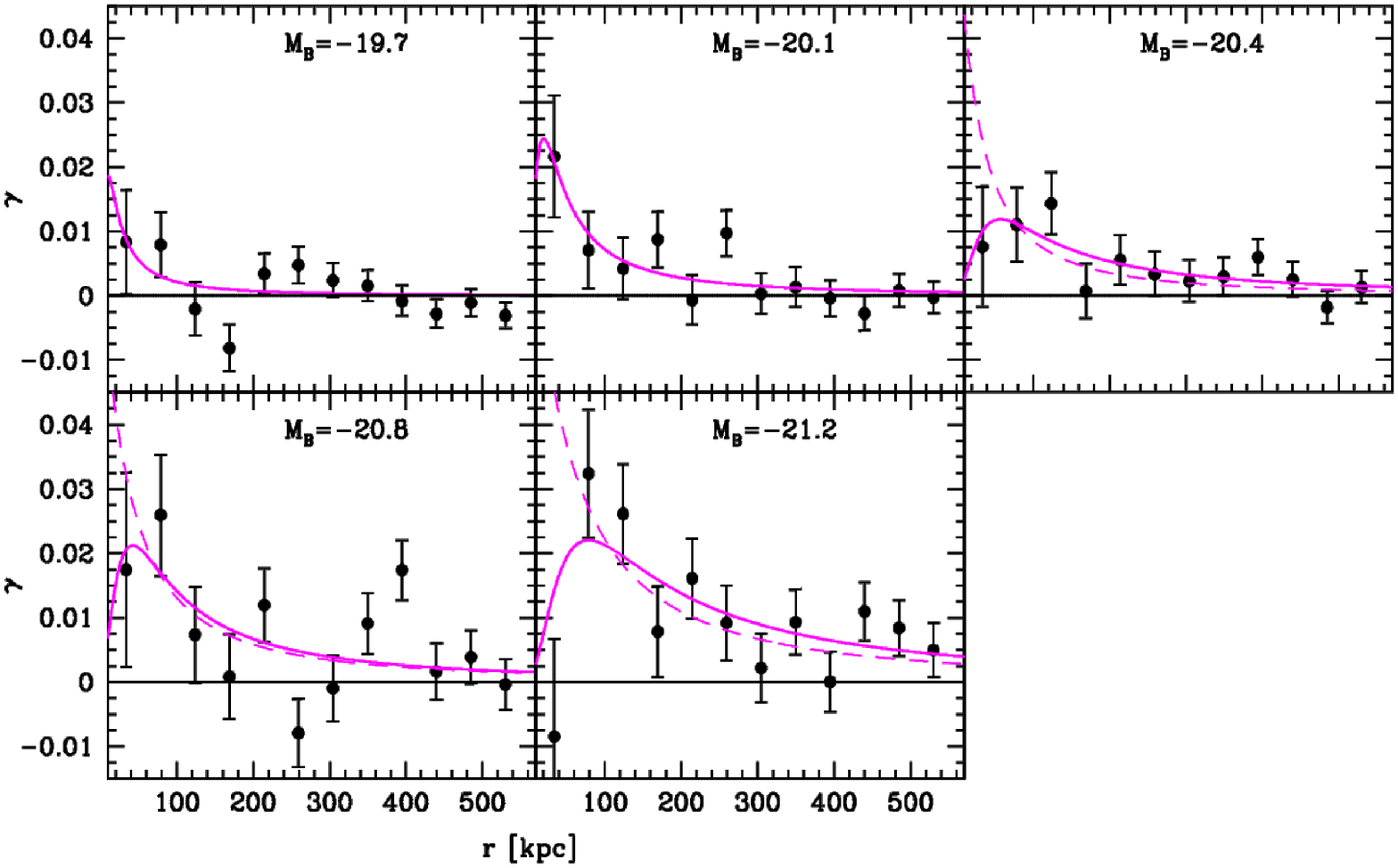,width=\textwidth}
\end{center}
\caption{Tangential shear measurements from Hoekstra et al. (2005)
as a function of projected distance from the lens in five R-band
luminosity bins. In this sample, the lenses are at a mean redshift
z$\sim$0.32 and the background sources are, in practice, at
$z=\infty$.  The solid (dashed) magenta line indicates the Burkert
(NFW) model fit to the data. At low luminosities they agree.}
\label{fig:weak_lensing}
\end{figure*}

\begin{table}
\begin{center}
\begin{tabular}{c|c|c|c}\hline
\emph{$M_{B}$}&\emph{$r_0\:$(kpc)}&\emph{$\rho_0 \:(10^{6}
M_{\odot}/kpc^3$)}&$\chi^2_{red}$\\ \hline
-19.7&$7^{+3}_{-6}$&$15^{+15}_{-7}$&1.6\\

-20.1&$14^{+6}_{-10}$&$10^{+10}_{-5}$&1\\

-20.4&$40.4^{+20}_{-20}$&$1.7^{+1.5}_{-0.7}$&0.7\\

-20.8&$30^{+10}_{-20}$&$4.1^{+4}_{-2}$&2.2\\

-21.1&$56^{+20}_{-20}$&$2.3^{+1.2}_{-0.6}$&1.1\\\hline
\end{tabular}
\caption{Structural parameters  and goodness of fit for a Burkert profile  to the  weak lensing signals of Hoekstra et al. (2005);
the corresponding  B magnitudes come  from their Table 1.}
\label{tab:weak_lensing}
\end{center}
\end{table}

Recent developments in weak gravitational lensing have made it
possible to probe the ensemble-averaged mass distribution around
galaxies out to large projected distances providing 
crucial information, complementary to that obtained from kinematics.
The tidal gravitational field of the DM halos generates weak-lensing
signals, by introducing small coherent distortions in the images of
distant background galaxies, which can be detected in current large
imaging surveys. We can measure, from the centre of the lenses out to
large distances (much greater than the distances probed by the
kinematic measurements), the azimuthally-averaged tangential shear
$\gamma_{\rm t}$
\begin{equation}
<\gamma_{\rm t}> \equiv
\frac{\overline{\Sigma}(R) - \Sigma(R)}{\Sigma_{\rm c}},
\end{equation}
where $\Sigma(R) = 2 \int_{0}^{\infty}\rho(R,z)dz$ is the projected
mass density of the object distorting the galaxy image, at projected
radius $R$ and $\overline{\Sigma}(R)= (2/R^2) \int_0^R x\Sigma(x) dx$
is the mean projected mass density interior to the radius $R$. The
critical density $\Sigma_{\rm c}$ is given by $ \Sigma_c \equiv
\frac{c^2}{4\pi G} \frac{D_{\rm s}}{D_{\rm l} D_{\rm ls}}$, where
$D_{\rm s}$ and $D_{\rm l}$ are the distances from the observer to the
source and lens, respectively, and $D_{\rm ls}$ is the source-lens
distance.  The above relations directly relate observed signals with
the underlying DM halo density. For our analysis we use the weak
lensing measurements from Hoekstra et al. (2005) available out to a
projected source-lens distance of $530$ kpc.  The sample, which
contains about $10^5$ isolated objects and spans the whole luminosity
range of Spirals, is split into 5 luminosity bins whose  B magnitudes (taken from their Table 1)
are given in Table~\ref{tab:weak_lensing}.  By adopting a density profile,
we model $\gamma_{\rm t}$ (see Figure~\ref{fig:weak_lensing}) and
obtain the structural parameters $\rho_0$ and $r_0$ by means of
standard best-fitting techniques.  The Burkert profile given by
equation~\ref{burkert} provides an excellent fit to the tangential
shear (see Figure~\ref{fig:weak_lensing} and
Table~\ref{tab:weak_lensing}).

Although testing the NFW density profile   is not an aim of this paper, let us  notice  that
it  provides a fit marginally sufficient  for the shear data, but  less  satisfactory  than
the Burkert profile especially around the most luminous objects ($M_B=-21.4$) 
(Figure~\ref{fig:weak_lensing}; see also Figure~6 of Hoekstra et al. 2005) we found $M_{\rm
vir} =4.2 \times  10^{12}$  where we found a reduced  $\chi^2$ of 2.  The region mapped by
weak lensing  is  much more extended  with respect to that probed by  internal  kinematics;
it is  therefore not surprising that  a NFW  halo  does not show   the same variance with
observations  found at  smaller radii  in that the  densities  of  actual  DM halos around
galaxies  seem to converge, for $R>1/3 R_{\rm vir}$  to NFW profile (see Salucci et al.
2008).
 
Notice that at low  luminosities ($M_B> -20.1$) the signal-to-noise is too low to
discriminate between mass models, so,  differently from  the other estimates  in this paper,
in these cases   we cannot  prove   a-posteriori  that the    Burkert profile  is superior 
over the cuspy one.    \end{document}